\documentclass[%
 reprint,
 superscriptaddress,
 amsmath,amssymb,
 aps,
]{revtex4-2}

\usepackage{graphicx}% Include figure files
\usepackage{dcolumn}% Align table columns on decimal point
\usepackage{bm}% bold math
\usepackage{hyperref}% add hypertext capabilities
\usepackage{color}%追加
\usepackage{comment}%追加
\newcommand{\Add}[1]{\textcolor{black}{#1}}	%←追加
\newcommand{\Addadd}[1]{\textcolor{black}{#1}}	%←追加
\newcommand{\Erase}[1]{\if0{#1}\fi}	%←追加
\newcommand{\Eraseadd}[1]{\if0{#1}\fi}	%←追加
\newcommand{\AddA}[1]{\textcolor{black}{#1}}
\newcommand{\EraseA}[1]{\if0{#1}\fi}	%←追加

\newcommand{\AddRevise}[1]{\textcolor{black}{#1}}	%←追加
\newcommand{\EraseRevise}[1]{\if0{#1}\fi}

\begin{document}

%\preprint{APS/123-QED}

\title{Single-shot high-resolution identification of discrete frequency modes of single-photon-level optical pulses}% Force line breaks with \\

\author{Daisuke Yoshida}
\email{yoshida-daisuke-cw@ynu.jp}
\affiliation{
Yokohama National University, 79-5 Tokiwadai, Hodogaya, Yokohama 240-8501, Japan
}
\affiliation{LQUOM Inc., 79-5 Tokiwadai, Hodogaya, Yokohama 240-8501, Japan}
\author{Mayuka Ichihara}%
\affiliation{
Yokohama National University, 79-5 Tokiwadai, Hodogaya, Yokohama 240-8501, Japan
}
\author{Takeshi Kondo}
\affiliation{
Yokohama National University, 79-5 Tokiwadai, Hodogaya, Yokohama 240-8501, Japan
}
\author{Feng-Lei Hong}
\affiliation{
Yokohama National University, 79-5 Tokiwadai, Hodogaya, Yokohama 240-8501, Japan
}
\author{Tomoyuki Horikiri}
\affiliation{
Yokohama National University, 79-5 Tokiwadai, Hodogaya, Yokohama 240-8501, Japan
}
\Erase{
\affiliation{
JST, PRESTO, 4-1-8 Honcho, Kawaguchi, Saitama, 332-0012, Japan
}
}
\date{\today}

\begin{abstract}
Frequency-multiplexed quantum communication usually requires a single-shot identification of the frequency mode of a single photon \Erase{unless a non-destructive measurement of the photon is possible}.
In this paper, we propose a scheme that can identify the frequency mode with high-resolution even for spontaneously emitted photons whose generation time is unknown, by combining the time-\Add{to-}space and frequency-\Add{to-}time mode mapping. 
We also demonstrate the mapping of the frequency mode (100 MHz intervals) to the \Erase{time}\Add{temporal} mode (435 ns intervals) for weak coherent pulses using atomic frequency combs.
\Add{This frequency interval is close to the minimum frequency mode interval of the \AddRevise{a}tomic frequency comb quantum memory with $\mathrm{Pr^{3+}}$ ion-doped $\mathrm{Y_2SiO_5}$ crystal, and the proposed scheme has the potential to maximize the frequency multiplexing of the quantum repeater scheme with the memory.}
 \Erase{This scheme engender\Add{may have} the potential for use in frequency multiplexed quantum communication. }
\end{abstract}

\maketitle

%\tableofcontents
%%%%%%%イントロ！！！！%%%%%%%
\section{\label{level1-1}INTRODUCTION}

The realization of quantum communication \Erase{using quantum states} enables various applications such as quantum key distribution~\cite{bb84,bbm92}, \Add{blind} quantum \Erase{cloud} computing~\cite{Broadbent2009}, \Addadd{and} atomic clocks with unprecedented stability and accuracy \cite{Komar2014}\Erase{, and so on}.
Quantum repeaters enable long distance quantum communication and are expected to constitute the core technology for the future quantum Internet~\cite{Kimble2008}. For these reasons, extensive \Add{studies}\Erase{research} ha\Add{ve}\Erase{s} been conducted in recent years toward their realization. 
One trend is frequency multiplexing \Add{ which is necessary for improving entanglement distribution rate} for quantum communication~\cite{Sinclair2014, Wengerowsky2018}.
In frequency-multiplexed quantum communication, the identification of frequency modes is usually necessary.
The frequency mode identification of single photons used in quantum communication is more difficult than that of classical light\EraseRevise{.M}\AddRevise{m}ainly because it must be measured in a single-shot, unless non-destructive measurement of the photon is possible.

Recently, an impactful study on quantum repeaters was reported by Lago-Rivera et al.~\cite{Lago-Rivera2021}.
\Add{T}\Erase{t}he authors demonstrated a quantum repeater scheme~\cite{Simon2007} that combines time-division multiplexed absorptive quantum memories based on \Erase{the} \Add{a}\Erase{A}tomic \Add{f}\Erase{F}requency \Add{c}\Erase{C}omb (AFC) scheme \cite{Afzelius2009} and photon-pair sources.
The AFC is a comb-shaped absorption \Erase{line}\Add{profile}.
It can be used not only as a time-multiplexing but also as a frequency multiplexing memory~\cite{Seri2019, Yang2018, Sinclair2014}. Here, we consider frequency multiplexing of the quantum repeater scheme~\cite{Lago-Rivera2021,Simon2007}.
In this case, AFCs are required to store photons of multiple frequency modes, and photon sources are required to generate photon pairs of frequency modes corresponding to those of AFCs. 
In Ref.~\cite{Lago-Rivera2021}, the AFC was tailored in inhomogeneous broadening of $\mathrm{^{3}H_{4}\leftrightarrow}$ $\mathrm{^{1}D_{2}}$ transition in $\mathrm{Pr^{3+}}$ ion-doped $\mathrm{Y_2SiO_5}$ crystal (Pr:YSO) by hole-burning technique~\cite{Nilsson2004}.
In the case of the AFC in Pr:YSO, the upper bandwidth limit of one frequency mode is about 4 MHz and the lower limit of the interval between frequency modes is about 100 MHz when used as a memory capable of reading out stored photons on demand~\cite{Ortu2022}. These limits are determined by the hyperfine level spacing in Pr:YSO~\cite{Ortu2022}.
The region where AFCs are created is limited to within the inhomogeneous broadening of Pr:YSO. Therefore, in the case of frequency-multiplexed quantum repeater using AFC\Addadd{s} in Pr:YSO, the upper limit of the entanglement distribution rate can be increased by making the frequency-mode interval as narrow as possible. As a frequency-multiplexed photon pair source, cavity-enhanced spontaneous parametric down-conversion (cSPDC)~\cite{Ou1999, Riel_nder2016, Niizeki2020}, which is also used in Ref.~\cite{Lago-Rivera2021}, would be promising.
The frequency mode interval of the photons generated by cSPDC depends on the free spectral range (FSR) of the cavity.
\AddRevise{In Ref.~\cite{Lago-Rivera2021}, where entanglement between quantum memories was generated by single-photon interference, the pump laser for cSPDC was continuous-wave (CW).}
In the case of CW excitation, the time at which photon pairs are generated is unknown.
To summarize the above, for frequency multiplexing of the quantum repeater scheme performed in Ref.~\cite{Lago-Rivera2021}, a method to identify high-resolution frequency modes of narrow linewidth photons without photon generation time information is desired.

A frequency-\Add{to-}time mode mapper (FTMM)~\cite{Saglamyurek2016,Saglamyurek2014, Avenhaus2009, Davis2017} and a frequency-\Add{to-}space mode mapper (FSMM)~\cite{Cheng2019,Casas2021} allow frequency mode identification of single photons. Typical examples of FTMMs are wavelength dispersion~\cite{Avenhaus2009} and chirped fiber Bragg gratings~\cite{Davis2017}.
In general, these are superior in that they are easy to implement but face difficulties in achieving resolution below GHz.
An FTMM using \Erase{atomic ensemble}\Add{AFCs} is also possible~\cite{Saglamyurek2016,Saglamyurek2014}.
In this method, the resolution is determined by the homogeneous broadening width and the energy level spacing used, and currently, high-resolution below GHz has been achieved~\cite{Saglamyurek2014}. 
However, the FTMM that can be adapted to photons with narrow linewidths below 10 MHz has not been realized.
\Erase{Another}\Add{A} common drawback of FTMM\Addadd{s} is that \Eraseadd{it}\Addadd{they} require\Eraseadd{s} information about the generation time of the photon for which the frequency mode is to be identified.
In contrast, an FSMM can identify the frequency mode without information about the photon generation time since frequency is identified by \Erase{space}\Add{spatial} mode. However, in general it is not easy to achieve a high-resolution of less than GHz.

In this paper, we propose a scheme that can identify high-resolution frequency modes of narrow linewidth photons without photon generation time information by combining a time-\Add{to-}space mode mapper (TSMM) and FTMMs.
We also demonstrate a high-resolution FTMMs using AFC, which can be applied to narrow linewidth photons.
The demonstrated FTMM is capable of identifying frequency modes in the 100 MHz interval, which is close to the \Erase{free spectral range}\Add{FSR} of recent cSPDC two-photon sources~\cite{Lago-Rivera2021, Niizeki2020} and the lower limit of frequency interval of AFCs in Pr:YSO.

%%%%%%%スキーム！！！！%%%%%%%
\section{\label{level1-2}SCHEME}
In this section, we initially propose a frequency mode identification scheme that combines FTMMs and \Addadd{a} TSMM\Eraseadd{s}. \Erase{Then the AFC, which can be an FTMM and a TSMM\\} \Erase{with the AFC will be described.}\Add{Next, We describe FTMMs using AFCs.}

%%%%周波数識別スキーム
\subsection{\label{level2-1}Frequency mode identification}
\begin{figure*}
    \centering
    \includegraphics[width = 16cm]{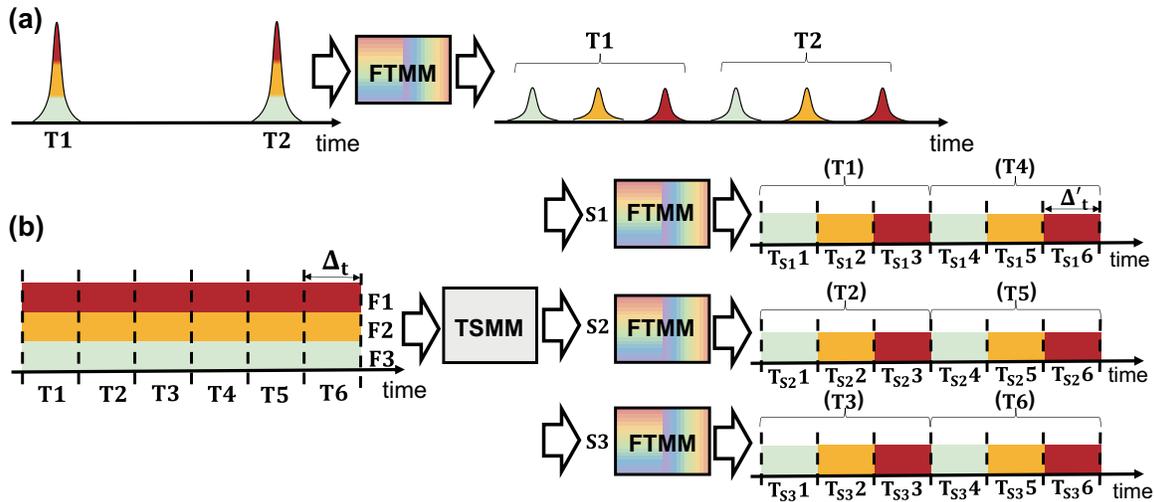}
    \caption{
    (a) Schematic of mapping frequency mode to \Erase{time}\Add{temporal} mode. 
    (b) Diagram of frequency mode identification combining a TSMM and FTMMs. 
    \Erase{time}\Add{temporal} mode T1 and T4 are mapped to \Erase{space}\Add{spatial} mode S1, \Erase{time}\Add{temporal} mode T2 and T4 are mapped to \Erase{space}\Add{spatial} mode S2, and \Erase{time}\Add{temporal} mode T3 and T6 are mapped to \Erase{space}\Add{spatial} mode \AddA{S3} using a TSMM. 
    The frequency mode is then mapped to \Erase{time}\Add{temporal} mode by an FTMM, which is provided for each \Erase{space}\Add{spatial} mode. 
    From the combination of \Erase{space}\Add{spatial} and \Erase{time}\Add{temporal} modes after mapping, the frequency mode can be uniquely determined.}
    \label{scheme}
\end{figure*}

First, we consider frequency-mode identification when only an FTMM is used.
A schematic diagram of this case is shown in Fig.~\ref{scheme}(a). 
By mapping multiple frequency modes, which exist in the same \Erase{time}\Add{temporal} mode \Erase{of different \Erase{time}\Add{temporal} modes}, it is possible to identify the frequency mode from the observation time of the photon if its generation time is known. 
In this case, however, to prevent multiple frequency modes from existing in the same \Erase{time}\Add{temporal} mode after frequency-\Add{to-}time mode mapping, the \Erase{\Erase{time}\Add{temporal} mode of the pulses}\Add{number of temporal modes} before mapping must be \Erase{thinned out}\Add{reduced}. From \Eraseadd{the communication rate point of view}\Addadd{the view of the communication rate}, \Erase{thinning out}\Add{reducing} the \Erase{\Erase{time}\Add{temporal}}\Add{number of temporal} modes is a disadvantage.
Therefore, we propose a frequency-mode identification scheme combining FTMMs and a TSMM.
In this scheme, frequency modes can be identified even if the photon generation time is unknown.
It also eliminates the need to \Erase{thwart}\Add{reduce} \Erase{\Erase{time}\Add{temporal}}\Add{the number of temporal} modes before FTMMs and a TSMM. 
The overview diagram is shown in Fig.~\ref{scheme}(b).
Consider the case where the generation time of the photon is unknown, i.e., the probability of the photon's existence is equally present at all times.
\Add{The duration of the temporal mode of  input photon is set to be $\mathit{\Delta_{\mathrm{t}}}$.}
\Erase{At this time,}
\Add{Here,}
$\mathit{\Delta_{\mathrm{t}}}$
is assumed to be sufficiently longer than the coherence time of the photon.
The TSMM converts each \Erase{time}\Add{temporal} mode separated by 
$\mathit{\Delta_{\mathrm{t}}}$
into \Erase{space}\Add{spatial} modes 
$\mathrm{S_{1}, S_{2},\cdots, S_{\it{N}_{\rm{S}}},S_{1},\cdots}$
w\Erase{h}ith $N_{\rm{S}}$ being the total number of \Erase{space}\Add{spatial} modes.
The \Erase{frequency-time mode mapper}\Add{FTMM} provided for each \Erase{space}\Add{spatial} mode converts each frequency mode to a \Erase{time}\Add{temporal} mode at $\mathit{\Delta_{\mathrm{t}}}^{\prime}$ intervals.
At this time, 
$\mathit{\Delta_{\mathrm{t}}}\leqq \mathit{\Delta_{\mathrm{t}}}^{\prime}$
must be satisfied to ensure that different frequency modes do not exist in the same \Erase{time}\Add{temporal} mode. For the same reason, if the total number of frequency modes is
$N_{\rm{F}}$, it must satisfy 
$N_\mathrm{F} \mathit{\Delta_{\mathrm{t}}^{\prime}} \leqq N_\mathrm{S} \mathit{\Delta_{\mathrm{t}}}$.
In this way, the frequency mode can be uniquely determined from the \Erase{space}\Add{spatial} and \Erase{time}\Add{temporal} modes in which the photon was observed.
Fig.~\ref{scheme}(b) shows the case 
$\mathit{\Delta_{\mathrm{t}} = \Delta_{\mathrm{t}}^{\prime}}$, $N_{\mathrm{F}} = N_{\mathrm{S}} = 3$.
The frequency resolution of this scheme depends on the resolution of the FTMMs.
A promising candidate for a high-resolution FTMMs is the AFC, which we describe below.
A promising candidate for the TSMM could be an optical switch array using  electro-optic modulators (EOMs)~\cite{Tu2019}.

\subsection{\label{level2-2}\Addadd{A} \Eraseadd{F}\Addadd{f}requency-to-time mode mapper \\using an atomic frequency comb}
\begin{figure}
    \centering
    \includegraphics[width = 8.5cm]{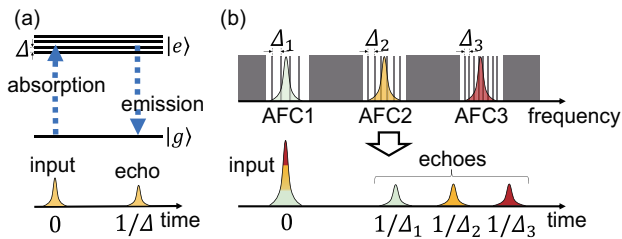}
    \caption{(a) Schematic diagram representing the function of AFC.
    The light pulse absorbed at time 0 is re-emitted at time 1/$\mathit{\Delta}$.
    (b) Schematic diagram of B with AFCs. AFCs with different comb spacing, $\mathit{\Delta_\mathrm{1}}$, $\mathit{\Delta_\mathrm{2}}$, $\mathit{\Delta_\mathrm{3}}$, are made for each frequency mode to be identified. 
    The light pulses are re-emitted in different \Erase{time}\Add{temporal} modes for each frequency mode.}
    \label{diag}
\end{figure}

The AFC is an equally spaced comb-\Erase{like}\Add{shaped} absorption \Erase{line}\Add{profile} as shown in Fig.~\ref{diag}(a).
Photons absorbed in the AFC are re-emitted in the same \Erase{space}\Add{spatial} mode in the inverse time of the comb spacing $\it{\Delta}$ (henceforth referred to as the echo signal).
Typically, AFCs are created by hole-burning in inhomogeneous broadening of rare-earth ion-doped crystals.
The AFC can also read out the absorbed photons on demand by using another ground level with no population~\cite{Afzelius2009, Afzelius2010, Timoney2013}.
In this study, frequency-\Add{to-}time mode \Erase{conversion}\Add{mapping} was performed by AFC, which is used as a fixed-time memory.

AFC is considered to be used as a high-resolution FTMM.
A schematic diagram of frequency-\Add{to-}time mode \Erase{conversion}\Add{mapping} using AFC is shown in Fig.~\ref{diag}(b).
Multiple AFCs can be tailored in inhomogeneous broadening~\cite{Seri2019, Yang2018, Sinclair2014}.
If AFCs are tailored with different comb intervals for each frequency mode, the time at which the echoes are reproduced can be changed for each frequency mode.
However, it is not possible to create multiple AFCs in arbitrary bands because creating a hole in one band will create \Erase{an} anti-hole\Add{s} in \Erase{an}other band\Add{s}.
\textcolor{black}{The lower limit of the interval between frequency modes is about 100 MHz when used as a memory capable of reading out stored photons on demand~\cite{Ortu2022}.}%<-イントロと同じことを言ってしまっているがどっちか削るべきか？
The spacing between adjacent AFCs should be at least ~100 MHz~\cite{Ortu2022}.
\Add{Therefore} \Erase{I}\Add{i}n this study, experiments were conducted to map three frequency modes separated by $\sim100$ MHz into \Erase{time}\Add{temporal} modes with $\mathit{\Delta_{\mathrm{t}}^{\prime}} \sim 435$ ns spacing. 

%%%%%%%実験系！！！！%%%%%%%
\section{\label{level1-3}EXPERIMENTAL SETUP}
\begin{figure*}
    \centering
    \includegraphics[width = 14cm]{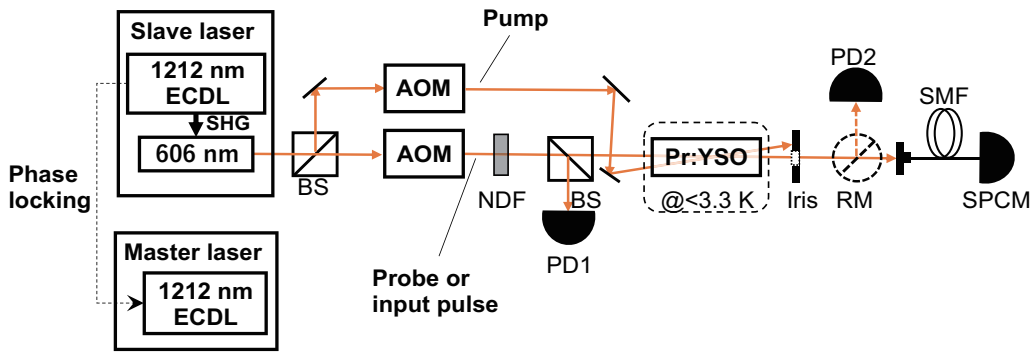}
    \caption{
    Experimental setup. 
    Second harmonic generation (SHG) of 1212 nm external cavity diode laser (ECDL) is used as pump, probe and input pulse\Addadd{s}.
    A neutral density filter (NDF) is used to reduce the input \Erase{light}\Add{pulse} to a single-photon-level. 
    The input \Erase{light}\Add{pulse} is coupled to a single mode fiber and observed by a SPCM.
    When observing AFC\Addadd{s}, the NDF is removed and the transmitted light was observed with a photo detector (PD1) instead of a SPCM. 
    The other photo detector (PD2) is used to monitor the probe laser intensity. 
    \Erase{HWP, half-wave plate; PM SMF, polarization maintaining\\} \Erase{single mode fiber; PBS, polarization beamsplitter;} BS, \Erase{non-polarization} beamsplitter\Add{; RM, removal mirror mount.}}
    \label{setup}
\end{figure*}

The overall experiment setup is illustrated in Fig.~\ref{setup}.
\Add{In our experiment, multiple AFCs were created within inhomogeneous broadening of Pr:YSO which had a $\mathrm{Pr^{3+}}$ doping rate of 0.05\% and dimensions of 3 mm × 3 mm × 5 mm.}
A wide modulation range is required for the laser beam to create an AFC in multiple \EraseA{bands}\AddA{frequency modes} by taking advantage of the inhomogeneous broadening of Pr:YSO (about 10 GHz~\cite{Nilsson2004}).
In this study, we \EraseRevise{created}\AddRevise{developed} a frequency stabilization and modulation system that can perform accurate GHz-order modulation 
\Add{\EraseRevise{by} using a dynamical phase lock technique~\cite{Numata2012}} and accurate and fast modulation of \EraseRevise{about}\AddRevise{approximately} 10 MHz \EraseRevise{by} using 
\Erase{a dynamical phase lock technique~\cite{Numata2012}, and} an acousto-optic modulator (AOM)~\cite{Donley2005}.
\Add{The absolute frequency of the master Laser (Toptica, TA pro) was stabilized to an optical frequency comb phase-locked to the GPS signals and the linewidth \EraseRevise{is}\AddRevise{was} narrowed using a reference cavity. 
The slave laser (Toptica, TA-SHG pro) used for pump, probe and input pulse\Addadd{s} was stabilized against \AddRevise{the} master laser.}
We used a closed-cycle cryostat (Montana instruments, cryostation) to cool the Pr:YSO crystal to $< 3.3$ K.
Each AFC was created by modulating the pump \Erase{light}\Add{laser} by ~10 MHz using the AOM in the double-path configuration~\cite{Donley2005}.
After each AFC is created, the laser itself is modulated by 100 MHz using the dynamic phase lock technique, and a different AFC was created by modulation with the AOM again.
The time required for 100 MHz modulation was \EraseRevise{about}\AddRevise{approximately} 10 ms in our setup. 
In this way, three AFCs with different comb spacing were created at 100 MHz intervals.

The probe laser for observing the created AFC can be turned on and off by \Add{an} AOM.
It was turned off during AFC creation and turned on only during observation.
During AFC observation, the probe laser was modulated by chirping the reference RF signal for phase locking in the dynamical phase lock technique~\cite{Numata2012}. 
The input pulse for observing the echo signal of the AFC was the same path as the probe \Erase{light}\Add{laser}. 
The input pulse was tailored by  \Add{the} AOM to be Gaussian with full width at half maximum (FWHM) of 5 MHz. The echo signal is coupled to a single mode fiber (SMF) and detected by a single photon counting module (SPCM, Perkin Elmer, \Erase{SPCM-CD 3254}\Add{SPCM-AQRH-14-FC}). %←万浪さん曰くこうらしいが、SPCM-AQRH-14-FCでは？？
The coupling efficiency of the SMF was \Erase{64}\Add{59}\%, the \Erase{\Add{(specification of the)}} detection efficiency \Erase{\Add{(at the corresponding efficiency)}} of the \textcolor{black}{SPCM was 59\%,} and the dark count rate is $\sim150$ Hz. %<=(検出効率datasheet 見て判断でよい？？)
\Erase{For intensity,}
the \Add{power of} pump \Erase{light}\Add{laser} was set to \Erase{$\sim1$ mW}\Add{$\sim2.6$ mW}, probe \Erase{light}\Add{laser} to \Erase{7}\Add{1} µW, and input pulse to mean photon number \Erase{$\mu = 5.1$}\Add{$\mu = 0.12$} per pulse. 
The beam diameter was set to $\sim500$ µm for the pump \Eraseadd{beam} and $\sim100$ µm for the probe and input \Eraseadd{beams}\Addadd{pulses}. These polarization\AddRevise{s} were aligned with the D2 axis of the Pr:YSO crystal.

%%%%%%%実験結果！！！！%%%%%%%
\section{\label{level1-4}EXPERIMENTAL RESULT}
\begin{figure}
    \centering
    \includegraphics[width = 8.5cm]{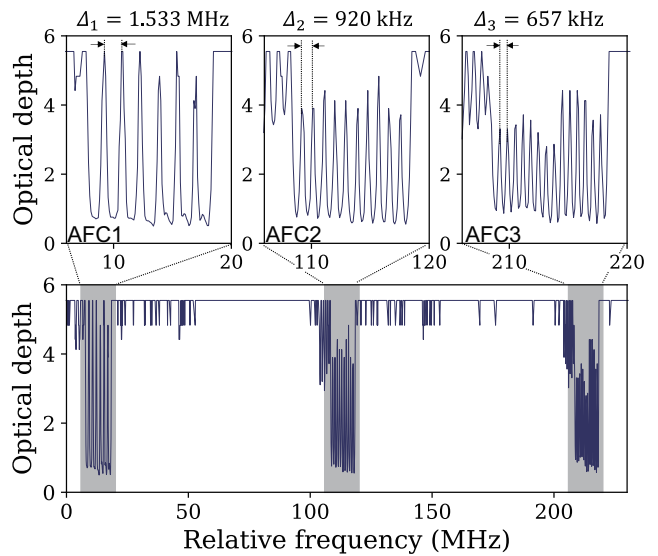}
    \caption{An overall view of the three AFCs and an enlarged view of each AFC.}
    \label{AFC_all}
\end{figure}

The three AFCs were observed by chirping a weak probe \Erase{light}\Add{laser} at a chirp rate of 5.2 MHz/ms.
The optical depth shown in Fig.~\ref{AFC_all} was the average of 16 measurements. It can be seen that AFCs with different comb spacing\AddRevise{s} were created.
In this experiment, the comb spacing was set to $\mathit{\Delta}_1 = 1.533$ MHz, $\mathit{\Delta}_2 = \Erase{952}\Add{920}$ kHz, and $\mathit{\Delta}_3 = 657$ kHz, respectively, in order from the low frequency side. 
This means that the expected echo times are $t_1 = 65\Erase{4}\Add{2}$ ns, $t_2 = 10\Erase{50}\Add{87}$ ns\AddA{, and $t_3 = 1522$ ns}.

\begin{figure}
    \centering
    \includegraphics[width = 8.5cm]{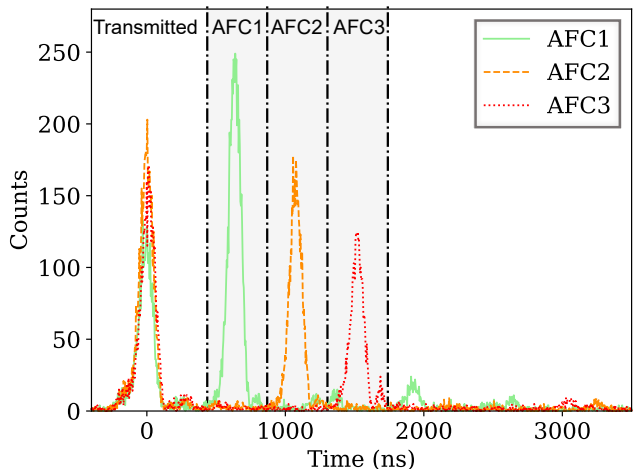}
    \caption{
    Echo signal observation results.
    \AddRevise{The bin size is 4.096 ns.}
    The grey shaded area surrounded by \AddRevise{dashed-dotted black} lines is the \Erase{time}\Add{temporal} mode in which each echo is expected to appear. 
    The second echo caused by reabsorption of the echo signal was also observed, and the third echo caused by AFC3 was sufficiently small and was considered negligible.}
    \label{echo}
\end{figure}

In this experiment, for each \AddA{frequency mode} created with multiple frequency modes, 1000\AddA{0} Gaussian pulses were input \Erase{for only one frequency mode}.
\Add{By repeating this procedure, }
\Add{t}he total number of pulses input for each frequency mode was $N_{\mathrm{in}}=\Erase{2.4}\Add{7.5}\times10^5$. The observed results are shown in Fig.~\ref{echo}.
It can be seen that the echoes appear in the expected \Erase{time}\Add{temporal} mode for each frequency mode.
Table~\ref{tab1} shows the expected \Erase{time}\Add{temporal} mode of detection for each frequency mode, echo efficiency $\eta_{\mathrm{echo}}$, and probability $\eta_{\mathrm{error}}$ of observing a photon in the other two expected \Erase{time}\Add{temporal} modes.
 $\eta_{\mathrm{echo}}$ is the ratio of the number of counts within the expected \Erase{time}\Add{temporal} mode (corrected for \Eraseadd{APD}\Addadd{SPCM} detection efficiency and fiber coupling loss) to the total number of photons input ($N_{\mathrm{in}} \mu$).
$\eta_{\mathrm{error}}$ is the ratio of the number of counts within the other two \Erase{time}\Add{temporal} modes that are not expected (corrected for SPCM detection efficiency and fiber coupling loss) to $N_{\mathrm{in}} \mu$.

%\begin{comment}
%\begin{table}[b]%The best place to locate the table environment is directly after its first reference in text
%\caption{\label{tab1}%
%Comb spacing \Erase{time}\Add{temporal} mode for each AFC, and experimental results for $\eta_{\rm{echo}}$ and %$\eta_{\rm{error}}$.
%}
%\begin{ruledtabular}
%\begin{tabular}{lccc}
%\multicolumn{1}{c}{\textrm{AFC}}&
%{\textrm{\Erase{time}\Add{temporal} mode}} (ns)&
%{$\eta_{\rm{echo}}$} (\%)&
%{$\eta_{\rm{error}}$} (\%)\\
%\colrule
%AFC1 ($\Delta_1=1.53$ MHz) & 437-870  & 13.5 & 2.1\\
%AFC2 ($\Delta_2=952$ kHz) & 870-1305 & 11.8 & 0.5\\
%AFC3 ($\Delta_3=657$ kHz) & 1305-1639 & 11.7 & 0.6\\
%end{tabular}
%\end{ruledtabular}
%\end{table}
%\end{comment}

\begin{table}[b]%The best place to locate the table environment is directly after its first reference in text
\caption{\label{tab1}%
Comb spacing and \Erase{time}\Add{temporal} mode for each AFC, and experimental results for $\eta_{\rm{echo}}$ and $\eta_{\rm{error}}$.\Add{Time window for each \Eraseadd{time}\Addadd{temporal} mode is 435 ns.}}
\begin{ruledtabular}
\begin{tabular}{lccc}
\multicolumn{1}{c}{\textrm{}}&
AFC1&
AFC2&
AFC3\\
\colrule
$\it{\Delta}$& 1.533 MHz  & \Erase{952}\Add{920} kHz & 657 kHz\\
\Erase{time}\Add{temporal} mode & \Erase{437-870}\Add{$652$} ns & \Erase{870-1305}\Add{1087} ns & \Erase{1305-1639}\Add{1522} ns\\
$\eta_{\rm{echo}}$ & 21\% & 14\% & 11\%\\
$\eta_{\rm{error}}$ & 2.2\%& 1.4\% & 1.2\%
\end{tabular}
\end{ruledtabular}
\end{table}
%%%%%%%議論！！！！%%%%%%%
\section{\label{level1-5}DISCUSSION}
In this \Erase{experiment}\Add{study}, three frequency modes at 100 MHz intervals were \Add{succesfully} mapped in different \Erase{time}\Add{temporal} modes. However, the probability of successful mapping is low, about 10\%.
Moreover, since the absorption efficiency of AFC is not unit, the \Add{photons that are transmitted without being absorbed}\Erase{transmitted \Erase{light}\Add{pulse}} exist\Erase{s} around 0 ns in Fig.~\ref{echo} and occup\Add{y}\Erase{ies} the \Erase{time}\Add{temporal} mode. Theoretically, near unit efficiency absorption and reemission can be obtained by using an AFC in a cavity~\cite{Afzelius2010_2}, which  would help \Add{re}solving these problems.

We consider the limits of the number of frequency modes for an FTMM using AFC\Add{s}. Factors that determine the limit include the modulation range of the \Add{pump} laser, inhomogeneous broadening, linewidth of the pump laser, and the creation time of the AFC.
In our system, the laser was directly modulated, but the modulation range is limited by the mode-hop range, which is about 15 GHz. Alternatively, the inhomogeneous width of the Pr:YSO doping rate of 0.05\% that we used is ~10 GHz~\cite{Nilsson2004}.
Therefore, if we were to create an AFC with different comb spacing every 100 MHz, the limit would be about 100 modes. In fact, if we assume a time width of 435 ns for one-\Erase{time}\Add{temporal} mode and try to create an AFC with different comb spacing in 100 modes, the comb spacing of the AFC with the smallest comb spacing will be ~20 kHz, and the linewidth of the pump laser must be sufficiently narrowed.
In the current system, the lower limit of the comb spacing that can be stably produced is about 500 kHz, and to achieve even smaller comb spacing, it is necessary to use ultra-low expansion cavity or self-heterodyne method for narrowing linewidth of the \Add{pump} laser~\cite{Young1999,Kefelian2009}.
The upper limit of the allowable AFC creation time is determined by the relaxation time between hyperfine levels, since the comb structure degrades after the AFC is created.
In this experiment, the time to create one AFC was $< 50$ ms and the time for frequency mode modulation was $\sim10$ ms.
Therefore, the time required to create an $N$ mode AFC\Addadd{s} is $< 60N$ ms. The extent to which the creation time is acceptable will depend on the system to which the AFC is applied.
\Eraseadd{The combination of this experimental system and a}
\Eraseadd{high-speed optical switch enables high-resolution}
\Eraseadd{frequency-mode identification of photons with an}
\Eraseadd{unknown photon generation time.}

This scheme is expected to be applied not only to multiplexing for quantum communication but also to various spectroscopic measurements. However, when AFC is used as an FTMM, frequency conversion is required to perform frequency mode identification of photons in various frequency bands.
\AddA{Coupling of \EraseRevise{a} AFC in Pr:YSO with telecommunication wavelength photons using frequency conversion are achieved~\cite{Maring2014,mannami2021}.}
\Erase{However}\Add{In contrast}, \Addadd{an FSMM}\Eraseadd{\Add{a} frequency-\Add{to-}space \Add{mode mapper}}\Erase{Aconversion} using VIPA~\cite{Casas2021} or gratings~\cite{Cheng2019} is also expected as a frequency\Erase{-} mode identification method. Compared to our scheme, they are superior in terms of ease of implementation \Add{and} frequency band extension\Erase{and configuration}.
The superiority of our scheme is that it allows for high-resolution frequency identification.
\Erase{\textcolor{black}{If the frequency-mode spacing to be identified is large enough,\\}} \Erase{there would be no advantage to employing our scheme.)} %<-要らんか？

%%%%%%%結論！！！！%%%%%%%
\section{\label{sec:level1}CONCLUSION}
In this study, we \EraseRevise{have} achieved frequency mode identification at 100 MHz intervals, which is close to the frequency multiplexing limit of AFC in Pr:YSO, a promising quantum memory for quantum repeater.
Since the FSR of a cSPDC source with high coupling to quantum memory has been demonstrated to be about 100 MHz~\cite{Lago-Rivera2021, Niizeki2020}, this spectroscopic system not only achieves the upper limit of discrete-mode spectral resolution with Pr:YSO, but also shows promising results for improving the quantum entanglement generation rate.

\begin{acknowledgments}
We thank Ippei Nakamura for the useful discussion.
This work was supported by SECOM foundation, JST PRESTO (JPMJPR1769), JST START (ST292008BN), JSPS KAKENHI Grant Number JP20H02652 and NEDO (JPNP14012). 
We also acknowledge the members of the Quantum Internet Task Force, which is a research consortium to realize the 
Quantum Internet, for comprehensive and interdisciplinary discussions of the Quantum Internet.
\end{acknowledgments}

\bibliography{main}% Produces the bibliography via BibTeX.

\end{document}